\documentclass{aa}

\usepackage{graphicx}
\usepackage{natbib}
\begin{document}

\title{On the relationship between auroral and nebular
oxygen line intensities in spectra of H\,{\sc ii} regions}

\author{Leonid S.~Pilyugin}
  \offprints{L.S.~Pilyugin}

   \institute{   Main Astronomical Observatory
                 of National Academy of Sciences of Ukraine,
                 27 Zabolotnogo str., 03680 Kiev, Ukraine
                 (pilyugin@mao.kiev.ua)}

\date{Received 14 Marth 2005 / accepted }

\abstract{
We investigate relationships between observed auroral and nebular oxygen 
line fluxes in spectra of H\,{\sc ii} regions. We find a relation that 
is metallicity-dependent at low metallicities, but  
becomes independent of metallicity (within the uncertainties of 
the available data) above 12+logO/H $\sim$ 8.25, 
i.e. there is one-to-one correspondence 
(the ff -- relation) between auroral and nebular oxygen line fluxes 
in spectra of high-metallicity H\,{\sc ii} regions. 
The ff -- relation allows one to estimate the flux in the auroral line from 
strong oxygen line measurements only.
This solves the problem of the electron temperature (and, consequently,
abundance) determination in high-metallicity H\,{\sc ii} regions.
The ff -- relation confirms the basic idea of the ``empirical'' method, 
proposed by \citet{pageletal79} a quarter of a century ago, that the oxygen 
abundance in H\,{\sc ii} region can be esimated  from the strong oxygen lines 
measurements only. 
\keywords{galaxies: ISM - (ISM) H\,{\sc ii} region: abundances}
}

\titlerunning{Relationship between auroral and nebular oxygen line fluxes}

\authorrunning{L.S.~Pilyugin}

\maketitle

\section{Introduction}

Accurate abundances in H\,{\sc ii} regions can be derived via the classic
T$_{e}$ -- method, T$_{e}$ being the electron temperature of the  H\,{\sc ii}
region. Measurements of faint auroral lines, such as
[O\,{\sc iii}]$\lambda$4363, are necessary to determine T$_{e}$.
Unfortunately, the auroral lines drop below detectability in the spectra
of high-metallicity H\,{\sc ii} regions. The electron temperature T$_e$ is 
an indicator of the physical conditions in H\,{\sc ii} regions.
\citet{lcal,hcal,m101,vybor} advocated that the physical conditions in an 
H\,{\sc ii} region can be estimated via the excitation parameter P, i.e. using 
the strong oxygen nebular lines only. If this is the case then a relationship 
between oxygen auroral and nebular line fluxes in spectra of H\,{\sc ii} 
regions must exist. This problem is examined empirically here.

We will be using the following notations throughout the paper:
R$_2$ = I$_{{\rm [OII] \lambda 3727+ \lambda 3729}}$/I$_{{\rm H\beta }}$,
R$_3$ = I$_{{\rm [OIII] \lambda 4959+ \lambda 5007}}$/I$_{{\rm H\beta }}$,
R = I$_{{\rm [OIII] \lambda 4363}}$/I$_{{\rm H\beta }}$, R$_{23}$ = R$_2$ + R$_3$.
With these definitions, the excitation parameter P can be 
expressed as: P = R$_3$/(R$_2$+R$_3$).

\section{Observational data: line fluxes}

We need to select a sample of H\,{\sc ii} regions which have oxygen
line fluxes (including the auroral line [O\,{\sc iii}]$\lambda 4363$) measured
with high precision.
The sample of low-metallicity H\,{\sc ii} regions in blue compact dwarf
galaxies observed by Izotov with colleagues
\citep{frickeetal01,gusevaetal01,gusevaetal03a,gusevaetal03b,
izotovthuan98,izotovthuan04,izotovetal94,izotovetal97,izotovetal01,
izotovetal04,noeskeetal00,thuanetal95,thuanetal99} 
meets that precision criterion. It constitutes one of the largest and most
homogeneous data sets now available, being obtained and reduced in the same way.
However, this sample contains only a few high-metallicity H\,{\sc ii} regions.
Additional recent measurements of the oxygen line fluxes in high-metallicity
(12+logO/H $>$ 8.25) H\,{\sc ii} regions were taken from
\citet{caplanetal00,estebanetal04,estebanetal05,
garciarojasetal04,kennicuttetal03,leeetal03,oeyshields00,
oeyetal00,peimbert03,tsamisetal03,vermeijetal02,vilcheziglesias03}. 
The spectroscopic data so assembled form the basis of the present study.

\section{Relationship}

\begin{figure}
\resizebox{1.00\hsize}{!}{\includegraphics[angle=000]{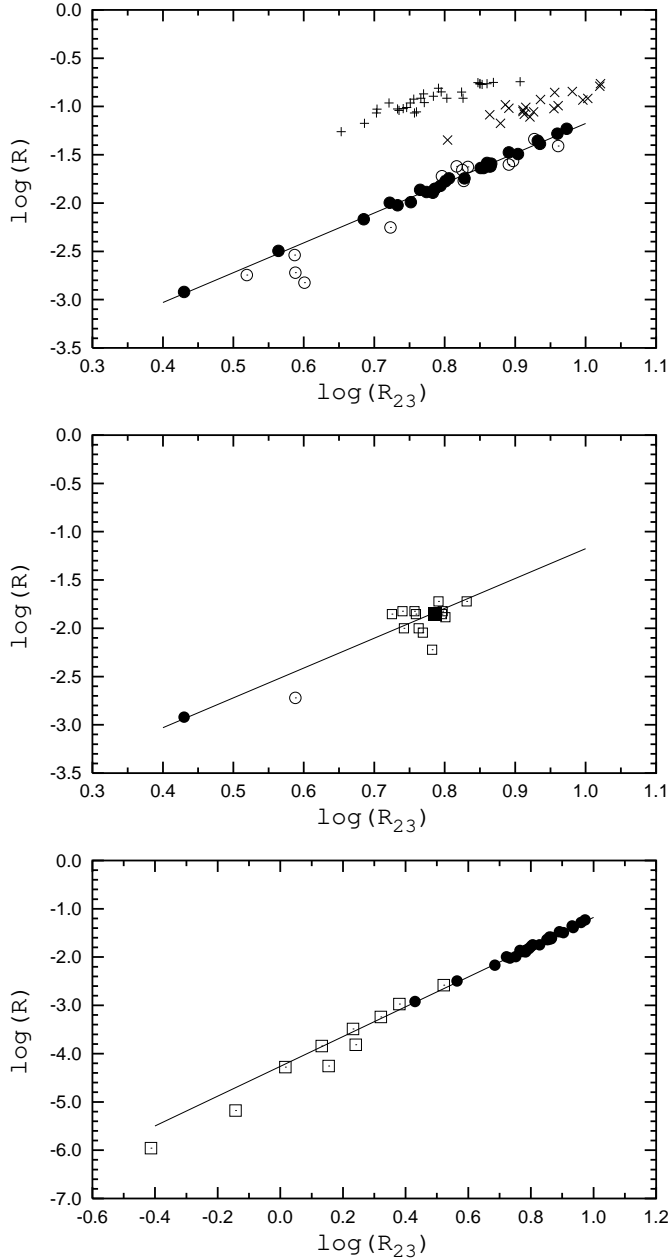}}
\caption{
{\it Top panel}. Flux in the oxygen auroral line R as a function of the 
total flux in strong nebular lines R$_{23}$. 
The crosses are H\,{\sc ii} regions with 7.8 $<$ 12+logO/H $<$ 8.0.
The plus signs are those with 7.4 $<$ 12+logO/H $<$ 7.6.
The circles (open + filled) are H\,{\sc ii} regions with 
12+logO/H $>$ 8.25, the filled circles are points used for 
determination of the ff -- relation (the solid line). 
{\it Middle panel}. The open squares are measurements for individual areas of 
the H\,{\sc ii} region DEM~323 in the Large Magellanic Cloud, the filled square 
is the integrated data \citep{oeyshields00,oeyetal00}. 
The filled circle is the measurement for the area of the H\,{\sc ii} region M~16 
in the Milky Way Galaxy from \citet{caplanetal00}, the open circle is 
that for the same H\,{\sc ii} region from \citet{estebanetal05}. 
The line is the same as in the top panel. 
{\it Bottom panel}. The open squares are faint 
H\,{\sc ii} regions from \citet{bresetal04}. The values of R for those 
H\,{\sc ii} regions are estimated from the electron temperatures 
(see text). The filled circles and the line are the same as in 
the top panel. 
}
\label{figure:f-f}
\end{figure}

The oxygen abundances in H\,{\sc ii} regions from our sample are in the range
from 12+logO/H $\sim$ 7.2 to $\sim$ 8.6.
The auroral line flux R is shown as a function of the total flux in strong
lines R$_{23}$ in Fig.~\ref{figure:f-f} (top panel).
The plus signs are H\,{\sc ii} regions with 7.4 $<$ 12+logO/H $<$ 7.6.
The crosses are those with 7.8 $<$ 12+logO/H $<$ 8.0. The circles (filled + 
open) are those with 12+logO/H $>$ 8.25. Inspection of Fig.~\ref{figure:f-f} 
shows that the flux in the auroral line R is linked to the total flux in 
the strong nebular lines R$_{23}$ through a relation of the type
\begin{equation}
\log R = a + b \times \log R_{23}.
\label{equation:ff}
\end{equation}
Eq.(\ref{equation:ff}) will be referred to hereinafter as the 
flux -- flux or ff -- relation. 
In the general case, the coefficient $a$ (and may be also $b$) 
in the ff -- relation is a function of the metallicity.

The top panel of Fig.~\ref{figure:f-f} shows that all the H\,{\sc ii} 
regions with 12+logO/H $>$ 8.25 lie along a single line, i.e. the 
coefficient $a$ becomes metallicity-independent at high metallicities
(or this dependence becomes very weak and is masked by errors in the flux 
measurements.) However, the scatter in the 
R -- R$_{23}$ diagram for high-metallicity objects is appreciable 
even for the recent measurements from the publications cited 
above. Therefore, the ff -- relation is derived by using an iteration 
procedure. In a first step, the relation is determined from all data
through the least squares method. Then, the point with  
largest deviation ($ \Delta$logR)$_{{\rm max}}$ 
is rejected, and a new relation is derived. The iteration procedure is 
stopped at the value of ($ \Delta$logR)$_{{\rm max}}$ = $\pm$0.05 
(the filled circles in the top panel of Fig.~\ref{figure:f-f}). 
The following values of the coefficients were obtained: $a$ = -- 4.264 $\pm$ 0.038 
and $b$ = 3.087 $\pm$ 0.046.
The ff -- relation is shown in Fig.~\ref{figure:f-f} by the line. 

Is the scatter in the R -- R$_{23}$ diagram real? If this is the case then it 
is necessary to establish the origin of the scatter, i.e. to establish 
a ``third'' parameter which is responsible for the scatter in the ff -- relation.
Small deviations of positions of H\,{\sc ii} regions from the ff -- relation 
can be naturally explained by the uncertainties in the flux measurements. 
It should be emphasized that the largest contribution to the 
deviation $\Delta$(logR) of the measured value of logR from the line does 
not necessarily come from errors in the measurement of R. In fact, a large 
error in the measurement (and/or in the dereddening) of 
the weak auroral line (say, 40\%) and much smaller (10\%) error of 
the strong nebular lines result in the same value of the 
discrepany $\Delta$(logR) $\sim$ 0.15 dex, since R $\sim$ R$_{23}^{3.087}$. 
Examination of the top panel of Fig.~\ref{figure:f-f} shows that three 
Galactic H\,{\sc ii} regions (M~16, M~20 and S~311 from the list of 
\citet{estebanetal05}) have large deviations from the line, of order 
0.2 dex.  It does not seem possible to explain this from errors in the 
measurements. 

There are other reasons why an H\,{\sc ii} region may give measurements 
lying off the line. 
It has been noted \citep{hcal} that a combination of the nebular line 
fluxes (the excitation parameter P) is a 
good indicator of the physical conditions in an H\,{\sc ii} region 
if two conditions are satisfied; 
{\it i}) the measured fluxes reflect their relative 
contributions to the radiation of the whole nebula, 
{\it ii}) the H\,{\sc ii} region is ionization-bounded. 
It can be suggested from general considerations that the ff -- relation 
also depends on these two conditions being satisfied. This suggests that 
the deviation from the ff -- relation can occur if the line fluxes 
measured within the slit do not reflect their relative contributions to the 
radiation of the nebula as a whole. This would happen, for example, if 
the relative fractions of the [O\,{\sc ii}] and [O\,{\sc iii}] zones in the 
slit differ significantly from those of the entire nebula. 
This effect would be more important for extended H\,{\sc ii} regions or 
H\,{\sc ii} regions with multiple emission peaks (due, e.g., to 
spatially distributed separate ionization sources) when 
only a small fraction of such an H\,{\sc ii} region is within the slit. 
This suggestion can be tested in the following way. 
The middle panel of Fig.~\ref{figure:f-f} shows the measurements for 
multiple positions in the H\,{\sc ii} region DEM~323 in the Large 
Magellanic Cloud (the open squares) and 
the integrated data (the filled square) from \citet{oeyshields00,oeyetal00}. 
The integrated data are in agreement with the ff -- relation while 
those for some individual areas show significant deviations. 
Comparison between integrated fluxes and 
those for individual areas shows that the largest deviation occurs 
when the relative contribution of R$_2$ flux to R$_{23}$ is 
significantly higher than in the integrated spectrum.  

There are two recent measurements of the oxygen line fluxes in the Galactic 
H\,{\sc ii} region M~16 \citep{caplanetal00,estebanetal05}. 
Its position based on the data of 
\citet{caplanetal00} (the filled circle in the middle panel of 
Fig.~\ref{figure:f-f}) is very close to the ff -- relation, 
whereas those of \citet{estebanetal05} 
(shown by the open circle) show a large deviation. 
Again, the relative contribution of the R$_2$ flux in the spectrum 
is significantly higher in the case of large deviation from the ff -- relation 
than that in the case of no (or small) deviation. 

\begin{figure}
\resizebox{1.00\hsize}{!}{\includegraphics[angle=000]{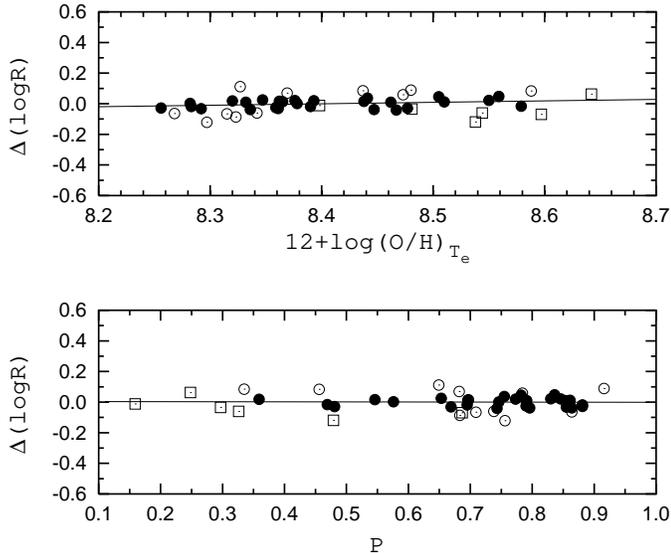}}
\caption{The residuals $\Delta$(logR) from the ff -- relation as a function of 
oxygen abundance ({\it top panel}) and of excitation parameter P 
({\it bottom panel}). 
The filled circles are points used for the determination of the ff -- relation, 
the lines are best fits to those data. 
The open circles are other H\,{\sc ii} regions with measured R. 
The open squares are H\,{\sc ii} regions 
from \citet{bresetal04} with R estimated. 
}
\label{figure:f-df}
\end{figure}

The ff -- relation was derived on the basis of bright H\,{\sc ii} regions. 
\citet{bresetal04} have recently detected the 
auroral line [S\,{\sc iii}]$\lambda$6312 and/or [N\,{\sc ii}]$\lambda$5755 
and determined the T$_e$-based abundances in 10 faint H\,{\sc ii} regions 
in the spiral galaxy M~51. Unfortunately they were unable to detect 
the oxygen 
auroral line R. We verify whether their faint H\,{\sc ii} regions follow 
the ff -- relation  in the following 
way. Using the electron temperature t$_3$ recommended by \citet{bresetal04}
and equations of the T$_e$-method, we have estimated the value of R which 
corresponds to the t$_3$ for every H\,{\sc ii} region. 
The bottom panel in Fig.~\ref{figure:f-f} shows the R versus R$_{23}$ diagram 
for the sample of faint H\,{\sc ii} regions from  \citet{bresetal04} 
(the open squares). The solid line is the ff -- relation from the top panel. 
The positions of faint H\,{\sc ii} regions in the R -- R$_{23}$ diagram 
show some kind of bifurcation. It is seen that the oxygen line fluxes in six 
faint H\,{\sc ii} regions (CCM~54, CCM~55, CCM~57, CCM~57A, CCM~71A, and 
CCM~84A) are in good agreement with the ff -- relation while the 
remaining four (CCM~10, CCM~53, CCM~72, and P~203) show significant deviations. 
The positions of those discrepant H\,{\sc ii} regions in 
the t$_{{\rm NII}}$ versus t$_{{\rm SIII}}$ diagram of \citet{bresetal04} (their Fig.2) 
suggest that their electron temperature determinations should be accurate, 
and, consequently, that their values of R estimated from the electron 
temperatures should be reliable. The deviation from the ff -- relation 
is thus probably due to uncertainties in the nebular 
oxygen line fluxes, and especially in the R$_2$ line flux since it 
makes a dominant contribution to 
the R$_{23}$ value, the flux in the R$_3$ line being small. 
 If this is the case, then the measured R$_2$ fluxes 
in the four H\,{\sc ii} regions CCM~10, CCM~53, CCM~72, 
and P~203 are overestimated by 20 to 30\%, 
higher than the errors of $\sim$ 10\% given by \citet{bresetal04}. 
As it was noted above the deviations from the ff -- relation can occur if 
the line fluxes measured within the slit do not reflect their relative 
contributions to the radiation of the nebula as a whole and 
this effect would be more important for extended  H\,{\sc ii} regions. 
It is interesting to note that at least three of the H\,{\sc ii} 
regions with large $\Delta$(logR)  (CCM~10, CCM~72, and P~203) 
have significantly larger angular diameters than the H\,{\sc ii} 
regions with small $\Delta$(logR)  in the sample.  \citet{bresetal04} 
have noted that large differences were found for CCM~10 and CCM~53 when 
they compare their line intensities with those reported by 
other authors. 

The residuals $\Delta$(logR) from the ff -- relation are shown in 
Fig.~\ref{figure:f-df} as a function of oxygen abundance (top panel) 
and of excitation parameter P (bottom panel). 
(H\,{\sc ii} regions with deviations greater than 
0.2 dex were excluded from consideration.)  
The filled circles are points used for the determination of the ff -- relation, 
the lines are best fits to those data. 
The open circles are other H\,{\sc ii} regions with measured R. 
The open squares are those with estimated R 
from the sample of \citet{bresetal04}. Fig.~\ref{figure:f-df} shows that 
the differences $ \Delta$(logR) do not show an appreciable correlation with 
metallicity or with the excitation parameter for the sample of objects with 
high-precision flux mesurements.

Thus there seems to be no ``third'' 
parameter in the ff -- relation, and the scatter 
in the R - R$_{23}$ diagram seems to be caused either by errors in the 
flux measurements or by the fact that the measured fluxes do not reflect 
their relative contributions to the radiation of the whole nebula. 
We need more high-precision data to confirm or reject this suggestion. 

\section{Conclusions}

A relationship between observed auroral and nebular oxygen line fluxes 
in spectra of H\,{\sc ii} regions was considered. It was 
found that this relationship is metallicity-dependent at low metallicities 
and becomes independent of metallicity (within the uncertainties of 
the available data) at metallicities higher than 12+logO/H $\sim$ 8.25, 
i.e. there is one-to-one correspondence 
(the ff -- relation) between auroral R and nebular oxygen line R$_{23}$ fluxes 
in spectra of high-metallicity H\,{\sc ii} regions
over more than three orders of magnitude in R.  

The ff -- relation allows one to estimate 
the auroral line flux from the measured nebular line fluxes. 
This solves the problem of the electron temperature (and, consequently,
abundance) determination in high-metallicity H\,{\sc ii} regions.

The ff -- relation confirms the basic idea of the ``empirical'' method, 
proposed by \citet{pageletal79} a quarter of a century ago, that the oxygen 
abundance in H\,{\sc ii} region can be esimated  from strong oxygen line 
measurements only. 

Among other things, the ff -- relation provides a possibility 
to select the H\,{\sc ii} regions with high quality line measurements 
and in which measured fluxes reflect their relative contributions to 
the radiation of the whole nebula. 

A detailed discussion of the application of the ff -- relation to abundance 
determinations in H\,{\sc ii} regions will be given elsewhere.

\begin{acknowledgements}
It is a pleasure to thank the referee, M.~Peimbert, for helpful
comments and suggestions.
I am grateful to B.E.J.~Pagel for constructive comments and suggestions as 
well as improving the English text.
I thank N.G.~Guseva, Y.I.~Izotov, T.X.~Thuan, and 
J.M.~ V\'{\i}lchez  for useful discussions.
This study was partly supported by grant No 02.07/00132 from the 
Ukrainian Fund of Fundamental Investigations. 
\end{acknowledgements}

\end{document}